\documentclass[aps,prd,superscriptaddress,nofootinbib,10pt,colorlinks,linkcolor=red,anchorcolor=green,citecolor=blue]{revtex4-2}

\usepackage{graphicx}
\usepackage{bm}
\usepackage{mathrsfs}
\usepackage{amsthm,amsmath,amssymb}
\usepackage{verbatim}
\usepackage{siunitx} 
\usepackage{booktabs} 
\usepackage{tensor}
\usepackage{braket}
\usepackage{orcidlink}
\usepackage{hyperref}

\begin{document}

\title{Fermions in external electromagnetic fields}

\author{Tianyang Li\,\orcidlink{0009-0008-5783-8353}}
\email{wyexlitianyang@tju.edu.cn}
\affiliation{Department of Physics, School of Science, Tianjin University, Tianjin 300354, China}

\author{Baoyi Chen\,\orcidlink{0000-0002-6411-2795}}
\email{baoyi.chen@tju.edu.cn}
\affiliation{Department of Physics, School of Science, Tianjin University, Tianjin 300354, China}
\affiliation{Department of Physics, Yantai University, Yantai 264005, China}

\author{Xuan Zhao\,\orcidlink{0009-0007-6093-5464}}
\email{zhaox21@mails.tsinghua.edu.cn}
\affiliation{Department of Physics, Tsinghua University, Beijing 100084, China}

\author{Pengfei Zhuang\,\orcidlink{0000-0002-9639-1493}}
\email{zhuangpf@mail.tsinghua.edu.cn}
\affiliation{Department of Physics, Yantai University, Yantai 264005, China}
\affiliation{Department of Physics, Tsinghua University, Beijing 100084, China}
\affiliation{Southern Center for Nuclear-Science Theory (SCNT), Institute of Modern Physics, Chinese Academy of Sciences, Huizhou 516000, China}

\begin{abstract}
We derive analytic solutions of the Dirac equation in external electromagnetic fields. For parallel electric and magnetic fields, orthonormal and complete in- and out-mode bases are constructed, and the corresponding causal in--out propagator is obtained. Exact solutions for perpendicular fields in the electrically dominated regime, $E>B$, are calculated by a Lorentz transformation. The method is further extended to general electric and magnetic fields with arbitrary relative orientation by combining a Lorentz boost with a spatial rotation. The solutions and in--out propagator provide the basis for studying dynamical processes in an external electromagnetic background.
\end{abstract}

\maketitle

\section{Introduction}
\label{sec:introduction}
Relativistic fermions in external electromagnetic fields are of continuing interest in quantum electrodynamics (QED), quantum chromodynamics (QCD), astrophysics, and high-energy nuclear physics. In relativistic heavy-ion collisions, extremely strong electromagnetic fields can be generated by the spectator charges and the induced currents in the quark--gluon plasma~\cite{Tuchin:2014hza,Voronyuk:2011jd,Deng:2012pc,Bzdak:2011yy}. The fields may affect a variety of phenomena, including the chiral magnetic effect~\cite{Kharzeev:2007jp,Fukushima:2008xe,Kharzeev:2015znc,Huang:2018wdl}, magnetic catalysis and inverse magnetic catalysis~\cite{Fukushima:2012kc,Mei:2024rjg,Yu:2014xoa,Fraga:2013ova,Cao:2021rwx}, spin polarization~\cite{Guo:2019joy}, and Schwinger pair production~\cite{Schwinger:1951nm,Kluger:1992gb,Hebenstreit:2010vz,Soldati:2011gi,Sheng:2018jwf,Sheng:2019ujr}. Although the electromagnetic fields created in realistic collisions are spatially inhomogeneous and time dependent, exact solutions in uniform backgrounds remain useful as analytical benchmarks and building blocks for locally constant-field approximations.

Exact solutions of the Dirac equation have been investigated for a variety of electromagnetic backgrounds. The solution in a uniform magnetic field has long been known and has found applications in strong-field QED and astrophysics~\cite{Kostenko:2018cgv,Kostenko:2019was}. Uniform electric fields have been studied in both time-dependent and static gauges~\cite{Nikishov:1969tt,Soldati:2011gi,Zhao:2025prd}. Solutions are also available for several spatially varying electric and magnetic fields, including exponential and Coulomb-type configurations~\cite{Gavrilov:2017ynq,Murguia:2009fs,PhysRevA.32.3251}. When electric and magnetic fields are simultaneously present, exact solutions have been obtained for mutually parallel fields~\cite{bagrov2014dirac}, perpendicular fields in the magnetically dominated regime~\cite{Canuto:1969ic}, and special configurations involving crossed fields and plane waves~\cite{narozhny1976solutions,bagrov2014dirac, PhysRevLett.98.116802}.

For strong-field quantum calculations, however, knowledge of the exact spinor solutions alone is not sufficient. The normalized in- and out-modes determine the external fermion states and their Bogoliubov transformation, thereby providing a direct description of vacuum pair production. The causal in--out propagator, on the other hand, describes the propagation of virtual fermions and is required for evaluating vacuum amplitudes and strong-field processes involving internal fermion lines. We should therefore not only construct exact stationary solutions and complete mode bases, but also derive the corresponding causal propagator in a form suitable for subsequent QED and QCD calculations.

In this work, we derive exact solutions of the Dirac equation for several representative electromagnetic-field configurations. For mutually parallel electric and magnetic fields, the equation is separated into a transverse harmonic-oscillator equation and a longitudinal Weber equation. The resulting stationary spinors are classified according to their asymptotic behavior, allowing us to construct orthonormal and complete in- and out-mode bases. We construct the corresponding Bogoliubov transformation and obtain the causal in--out propagator as an explicit mode sum. These results provide the external-state wave functions and fermion Green function needed for investigating vacuum pair production and other quantum processes in electromagnetic backgrounds. For mutually perpendicular fields in the electrically dominated regime, $E>B$, the exact laboratory-frame solutions are obtained by transforming the pure-electric-field solutions through a Lorentz boost. We then extend this construction to general electric and magnetic fields with an arbitrary relative orientation by combining a Lorentz boost, which renders the fields parallel, with a spatial rotation that aligns them along a common coordinate axis.

The paper is organized as follows. In Sec.~\ref{subsec:Solution-in-parallel} we derive the stationary solutions for parallel electric and magnetic fields, and in Sec.~\ref{sec:in-out-modes-and-propagator} we construct the in- and out-mode bases and the associated causal propagator, and demonstrate that its mode-sum expression reproduces the Schwinger proper-time representation. The perpendicular configuration with $E>B$ is considered in Sec.~\ref{sec:perpendicular-fields}, and the construction for a general field configuration is presented in Sec.~\ref{sec:arbitrary-fields}. We summarize in Sec.~\ref{sec:conclusions}. Details concerning the special functions and the normalization, orthogonality, and completeness of the stationary modes are provided in the appendices. Throughout the paper, we adopt the natural units with $\hbar=c=1$ and the metric convention $g_{\mu \nu}= \mathrm{diag}(+,-,-,-)$. The coordinate four-vector is denoted as $x^\mu=(t,{\bm x})=(t,x,y,z)$, and the symbol $e$ denotes the charge appearing in the gauge-covariant derivative. For definiteness, we take $e>0$, $E>0$, and $B>0$. We use the Dirac representation for the gamma matrices.

\section{Parallel electric and magnetic fields}
\label{subsec:Solution-in-parallel}
The Dirac equation for a fermion of mass $m$ and charge $e$ in an external electromagnetic field is
\begin{equation}
	\left(i\gamma^{\mu}D_{\mu}-m\right)\psi(t,\bm{x})=0,
\end{equation}
where $D_{\mu}=\partial_{\mu}+ieA_{\mu}(t,\bm{x})$ is the gauge-covariant derivative, and $A_{\mu}(t,\bm{x})$ denotes the external electromagnetic four-potential, whose specific form depends on the field configuration and gauge choice. In this section, we consider mutually parallel and uniform electric and magnetic fields oriented along the $-z$ direction. A suitable choice for the four-potential is $A_\mu(t,\bm{x})=(Ez,0,Bx,0)$. The corresponding Hamiltonian is 
\begin{equation}
	\hat{H}=-i\gamma^{0}\left[\gamma^{1}\partial_{x}+\gamma^{2}\left(\partial_{y}+ieBx\right)+\gamma^{3}\partial_{z}\right]+eEz+m\gamma^{0}.\label{eq:hamiltonian}
\end{equation}

Because the Hamiltonian is time-independent and invariant under a translation along the $y$ axis, the fermion energy $p_{0}$ and the momentum $p_{y}$ are strictly conserved quantities. Consequently, the stationary-state wave function can be factorized as
\begin{equation}
	\psi(t,\bm{x})=\mathrm{e}^{-i\left(p_{0}t-p_{y}y\right)}\phi(x,z),
    \label{eq:stationary-ansatz}
\end{equation}
and the Dirac equation is reduced to
\begin{equation}
	\left[\left(p_{0}-eEz\right)\gamma^{0}+i\gamma^{1}\partial_{x}-\left(p_{y}+eBx\right)\gamma^{2}+i\gamma^{3}\partial_{z}-m\right]\phi(x,z)=0.
\label{eq:first-order-parallel}
\end{equation}

To separate the spin and spatial parts of the wave function, we introduce the differential operator
\begin{equation}
	\mathcal{ P} = \left(p_{0}-eEz\right)\gamma^{0}+i\gamma^{1}\partial_{x}-\left(p_{y}+eBx\right)\gamma^{2}+i\gamma^{3}\partial_{z},
\label{eq:P-operator-parallel}
\end{equation}
and take the factorization 
\begin{equation}
	\phi(x,z) = (\mathcal{{P}}+m)f(x,z)\chi,
\label{eq:squared-ansatz-parallel}
\end{equation}
where $f(x,z)$ is a scalar function and $\chi$ is a constant bispinor. Using the relation
\begin{equation}
(\mathcal{{P}}-m)(\mathcal{{P}}+m)=\mathcal{{P}}^{2}-m^{2},
\end{equation}
we obtain
\begin{equation}
	\left[\partial^2_x+\partial^2_z+\left(p_{0}-eEz\right)^{2}-\left(p_{y}+eBx\right)^{2}-m^{2}+ieE\gamma^{0}\gamma^{3}-ieB\gamma^{1}\gamma^{2}\right]f(x,z)\chi=0.
\end{equation}
Observing that the matrix $\gamma^{0}\gamma^{3}$ and the spin component $\Sigma_z=i\gamma^1\gamma^2/2$ commute, we can select the four independent bispinors $\chi_{\epsilon s}$ ($\epsilon=\pm,s=\pm$) as their common eigenstates. This choice completely separates the coordinate degrees of freedom from the spin ones. In the Dirac representation of the $\gamma$ matrices, the specific forms of $\chi_{\epsilon s}$ are given by
\begin{equation}
	\chi_{++}=\frac{1}{\sqrt{2}}\begin{pmatrix}1\\0\\1\\0\end{pmatrix},\quad
	\chi_{+-}=\frac{1}{\sqrt{2}}\begin{pmatrix}0\\1\\0\\-1\end{pmatrix},\quad
	\chi_{-+}=\frac{1}{\sqrt{2}}\begin{pmatrix}1\\0\\-1\\0\end{pmatrix},\quad
	\chi_{--}=\frac{1}{\sqrt{2}}\begin{pmatrix}0\\1\\0\\1\end{pmatrix},
\end{equation}
which satisfy the eigenvalue equations,
\begin{equation}
	\Sigma_z\chi_{\epsilon s}=\frac{s}{2}\chi_{\epsilon s},\qquad \gamma^0\gamma^3\chi_{\epsilon s}=\epsilon\chi_{\epsilon s}
\end{equation}
and the normalization, orthogonality and completeness relations,
\begin{equation}
    \chi_{\epsilon s}^\dagger\chi_{\epsilon' s'} = \delta_{ss'}\delta_{\epsilon\epsilon'},\qquad
    \sum_{\epsilon=\pm}\sum_{s=\pm}\chi_{\epsilon s} \chi_{\epsilon s}^\dagger = I_4.
\end{equation}
Taking these bispinors, the scalar function $f_{\epsilon s}(x,z)$ consequently satisfies
\begin{equation}
	\left[\partial^2_x+\partial^2_z+\left(p_{0}-eEz\right)^{2}-\left(p_{y}+eBx\right)^{2}-m^{2}+ieE\epsilon-eBs\right]f_{\epsilon s}(x,z)=0.
\label{eq:scalar-parallel-dimensional}
\end{equation}

Introducing the dimensionless variables to replace the coordinates $x$ and $z$,
\begin{equation}
	\xi_{1} = \sqrt{eB}\left(x+\frac{p_y}{eB}\right),\qquad \xi_{2}	= \sqrt{eE} \left(z-\frac{p_0}{eE}\right),
\label{eq:parallel-dimensionless-variables}
\end{equation}
Eq.~\eqref{eq:scalar-parallel-dimensional} becomes
\begin{equation}
	\left[eB\frac{\partial^{2}}{\partial\xi_{1}^{2}}+eE\frac{\partial^{2}}{\partial\xi_{2}^{2}}-eB\xi_{1}^{2}+eE\xi_{2}^{2}-m^{2}+ieE\epsilon-eBs\right]f_{\epsilon s}(\xi_1,\xi_2)=0,
\label{scalar function equation}
\end{equation}
which allows us to separate the transverse and longitudinal coordinates $\xi_1$ and $\xi_2$ ($x$ and $z$),  
\begin{equation}
	f_{\epsilon s}(\xi_1,\xi_2) = g_s(\xi_1)h_\epsilon(\xi_2),
\end{equation}
and obtain two ordinary differential equations controlling the transverse and longitudinal motion of the fermion, 
\begin{eqnarray}
&& \left[eB\frac{d^{2}}{d\xi_{1}^{2}}-eB\xi_{1}^{2}-eBs+\kappa\right]g_{s}(\xi_{1})=0,\nonumber\\
&& \left[eE\frac{d^{2}}{d\xi_{2}^{2}}+eE\xi_{2}^{2}-m^{2}+ieE\epsilon-\kappa\right]h_{\epsilon}(\xi_{2})=0,
\label{eq:electric}
\end{eqnarray}
where $\kappa$ is a constant. The first equation is the harmonic oscillator equation, and the second one takes the form of the Weber equation~\cite{gradshteyn2014table}. The independent solutions can be chosen as 
\begin{eqnarray}
&& g_1(\xi_{1}) = \varphi_{n-1}(\xi_{1}),\nonumber\\
&& g_2(\xi_{1}) = \varphi_{n}(\xi_{1}),\nonumber\\
&& h_1(\xi_{2}\mid\pm) = D_{i\mu_{n}/2}\left(\pm(1+i)\xi_{2}\right),\nonumber\\
&& h_2(\xi_{2}\mid\pm) = D_{i\mu_{n}/2-1}\left(\pm(1+i)\xi_{2}\right),
\end{eqnarray}
where $\varphi_n(\xi_1)$ denotes the normalized harmonic oscillator function, $D_{\nu}(z)$ is the parabolic cylinder function, and $n=0,1,\cdots$ is the quantum number controlling the fermion transverse momentum. We have defined here the dimensionless parameter $\mu_{n}=(m^{2}+2neB)/(eE)$ and adopted the convention $\varphi_{-1}(\xi_1)=0$. Consequently, the independent solutions of the scalar equation $f_{\epsilon s}$ can be written as 
\begin{eqnarray}
		f_{++}(\xi_{1},\xi_{2}\mid\pm) &=& \varphi_{n-1}(\xi_{1}) D_{i\mu_{n}/2}\left(\pm(1+i)\xi_{2}\right),\nonumber\\
		f_{-+}(\xi_{1},\xi_{2}\mid\pm) &=& \varphi_{n-1}(\xi_{1}) D_{i\mu_{n}/2-1}\left(\pm(1+i)\xi_{2}\right),\nonumber\\
		f_{+-}(\xi_{1},\xi_{2}\mid\pm) &=& \varphi_{n}(\xi_{1}) D_{i\mu_{n}/2}\left(\pm(1+i)\xi_{2}\right),\nonumber\\
		f_{--}(\xi_{1},\xi_{2}\mid\pm) &=& \varphi_{n}(\xi_{1}) D_{i\mu_{n}/2-1}\left(\pm(1+i)\xi_{2}\right).
\end{eqnarray}

Taking into account the relations for the bispinors,
\begin{equation}
    \gamma^0\chi_{\epsilon s}=\chi_{(-\epsilon)s},\qquad \gamma^1\chi_{\epsilon s}=\epsilon s\chi_{\epsilon(-s)},\qquad \gamma^2\chi_{\epsilon s}=i\epsilon\chi_{\epsilon(-s)},\qquad \gamma^3\chi_{\epsilon s}=\epsilon\chi_{(-\epsilon)s}
\end{equation}
and the recurrence relations for the parabolic cylinder function and the harmonic oscillator function, see the Appendices~\ref{app:cylinder} and \ref{app:harmonic-oscillator}, we obtain the four wave functions $\phi_{sn}$ at fixed $\epsilon=-$,
\begin{eqnarray}
	\phi_{+n}(x,z\mid\pm) &=& (\mathcal{{P}}+m)f_{-+}(\xi_{1},\xi_{2}\mid\pm)\chi_{-+}\nonumber\\
	&=& \pm(i-1)\sqrt{eE}\,\varphi_{n-1}\left(\xi_{1}\right)D_{i\mu_{n}/2}\left(\pm(1+i)\xi_{2}\right)\chi_{++}\nonumber\\
	&& + i\sqrt{2neB}\,\varphi_{n}\left(\xi_{1}\right)D_{i\mu_{n}/2-1}\left(\pm(1+i)\xi_{2}\right)\chi_{--}
	   + m\varphi_{n-1}\left(\xi_{1}\right)D_{i\mu_{n}/2-1}\left(\pm(1+i)\xi_{2}\right)\chi_{-+},\nonumber\\
	\phi_{-n}(x,z\mid\pm) &=& (\mathcal{{P}}+m)f_{--}(\xi_{1},\xi_{2}\mid\pm)\chi_{--}\nonumber\\
	&=& \pm(i-1)\sqrt{eE}\,\varphi_{n}\left(\xi_{1}\right)D_{i\mu_{n}/2}\left(\pm(1+i)\xi_{2}\right)\chi_{+-}\nonumber\\
	&&+ i\sqrt{2neB}\,\varphi_{n-1}\left(\xi_{1}\right)D_{i\mu_{n}/2-1}\left(\pm(1+i)\xi_{2}\right)\chi_{-+}
	  + m\varphi_{n}\left(\xi_{1}\right)D_{i\mu_{n}/2-1}\left(\pm(1+i)\xi_{2}\right)\chi_{--}.
\label{eq:parallel}
\end{eqnarray}
The other four wave functions $(\mathcal{{P}}+m)f_{+s}(\xi_{1},\xi_{2}\mid\pm)\chi_{+s}$ at fixed $\epsilon=+$ are linear combinations of $\phi_{sn}$ with $\epsilon=-$ and therefore do not provide additional independent solutions of the first-order Dirac equation. Together with the plane-wave factor $\mathrm{e}^{-i\left(p_{0}t-p_{y}y\right)}$, the wave functions retained above constitute a convenient independent set of stationary modes for the parallel-field configuration. The normalization and asymptotic classification of these solutions will be specified below in constructing the in- and out-mode bases and the corresponding causal propagator.

\section{In- and out-mode bases and causal propagator}
\label{sec:in-out-modes-and-propagator}
\subsection{In- and out-modes and the mode-sum propagator}
We now classify the stationary solutions obtained above according to their asymptotic behavior and use them to construct the in- and out-mode bases. The four asymptotically defined stationary modes are
\begin{eqnarray}
        \psi_{sn}^{(1)}(t,\bm{x}) &=& (eB)^{1/4}/\sqrt{2\pi}\ C_n \mathrm{e}^{-i\left(p_{0}t-p_{y}y\right)}\phi_{sn}(x,z\mid+)\nonumber\\
        &=& (eB)^{1/4}/\sqrt{2\pi}\ C_n \mathrm{e}^{-i\left(p_{0}t-p_{y}y\right)}\left[A_1(z)U_{sn}(x)+B_1(z)V_{sn}(x)\right],\nonumber\\
        \psi_{sn}^{(2)}(t,\bm{x}) &=& (eB)^{1/4}/\sqrt{2\pi}\ C_n \mathrm{e}^{-i\left(p_{0}t-p_{y}y\right)}\phi_{sn}(x,z\mid-)\nonumber\\
        &=& (eB)^{1/4}/\sqrt{2\pi}\ C_n \mathrm{e}^{-i\left(p_{0}t-p_{y}y\right)}\left[A_2(z)U_{sn}(x)+B_2(z)V_{sn}(x)\right],\nonumber\\
        \psi_{sn}^{(3)}(t,\bm{x}) &=& (eB)^{1/4}/\sqrt{2\pi}\ D_n \mathrm{e}^{-i\left(p_{0}t-p_{y}y\right)}\left[\phi_{sn}(x,z\mid-)+\mathrm{e}^{\frac{\pi}{2}\mu_n}\phi_{sn}(x,z\mid+)\right]\nonumber \\ 
        &=& (eB)^{1/4}/\sqrt{2\pi}\ D_n \mathrm{e}^{-i\left(p_{0}t-p_{y}y\right)}\left[A_3(z)U_{sn}(x)+B_3(z)V_{sn}(x)\right],\nonumber\\
        \psi_{sn}^{(4)}(t,\bm{x})  &=& (eB)^{1/4}/\sqrt{2\pi}\ D_n \mathrm{e}^{-i\left(p_{0}t-p_{y}y\right)}\left[\phi_{sn}(x,z\mid+)+\mathrm{e}^{\frac{\pi}{2}\mu_n}\phi_{sn}(x,z\mid-)\right]\nonumber \\   
        &=& (eB)^{1/4}/\sqrt{2\pi}\ D_n \mathrm{e}^{-i\left(p_{0}t-p_{y}y\right)}\left[A_4(z)U_{sn}(x)+B_4(z)V_{sn}(x)\right]
\label{asymptotically defined stationary modes}
\end{eqnarray} 
with the normalization constants
\begin{equation}
    C_n = \frac{\mathrm{e}^{-3\pi\mu_n/8}}{\sqrt{4\pi eE}}, \qquad D_n = \frac{\mathrm{e}^{-3\pi\mu_n/8}}{\sqrt{4\pi e E}\sqrt{\mathrm{e}^{\pi\mu_n}-1}},
\end{equation}
the transverse spinors
\begin{eqnarray}
    U_{+n}(x) &=& \varphi_{n-1}\left(\xi_{1}\right)\chi_{++},\nonumber\\
    U_{-n}(x) &=& \varphi_{n}\left(\xi_{1}\right)\chi_{+-},\nonumber\\
    V_{+n}(x) &=& i\sqrt{2neB}\varphi_{n}\left(\xi_{1}\right)\chi_{--}+ m\varphi_{n-1}\left(\xi_{1}\right)\chi_{-+},\nonumber\\
    V_{-n}(x) &=& i\sqrt{2neB}\varphi_{n-1}\left(\xi_{1}\right)\chi_{-+}+m\varphi_{n}\left(\xi_{1}\right)\chi_{--}
    \label{eq:UandV}
\end{eqnarray}
and the longitudinal components
\begin{eqnarray}
A_1(z) &=& (i-1)\sqrt{eE}D_{i\mu_n/2}\left((1+i)\xi_{2}\right),\nonumber\\
B_1(z) &=& D_{i\mu_n/2-1}\left((1+i)\xi_{2}\right),\nonumber\\
A_2(z) &=& -(i-1)\sqrt{eE}D_{i\mu_n/2}\left(-(1+i)\xi_{2}\right),\nonumber\\
B_2(z) &=& D_{i\mu_n/2-1}\left(-(1+i)\xi_{2}\right),\nonumber\\
A_3(z) &=& (i-1)\sqrt{eE}\frac{i\sqrt{2\pi}\mathrm{e}^{\pi\mu_n/4}}{\Gamma(-i\mu_n/2)}D_{-i\mu_n/2-1}\left((1-i)\xi_{2}\right),\nonumber\\
B_3(z) &=& \frac{\sqrt{2\pi}\mathrm{e}^{\pi\mu_n/4}}{\Gamma(1-i\mu_n/2)}D_{-i\mu_n/2}\left((1-i)\xi_{2}\right),\nonumber\\
A_4(z) &=& (i-1)\sqrt{eE}\frac{-i\sqrt{2\pi}\mathrm{e}^{\pi\mu_n/4}}{\Gamma(-i\mu_n/2)}D_{-i\mu_n/2-1}\left(-(1-i)\xi_{2}\right),\nonumber\\
B_4(z) &=& \frac{\sqrt{2\pi}\mathrm{e}^{\pi\mu_n/4}}{\Gamma(1-i\mu_n/2)}D_{-i\mu_n/2}\left(-(1-i)\xi_{2}\right).
\end{eqnarray}
The label $s=\pm$ distinguishes the spinor modes used here and does not directly specify their conserved magnetic polarization. A basis of definite magnetic polarization can be constructed by taking suitable linear combinations of these modes. The corresponding conserved operator and its eigenstates are presented in Appendix~\ref{app:magnetic-polarization}. The stationary modes defined in Eq.~\eqref{asymptotically defined stationary modes} are selected according to their asymptotic behavior in the longitudinal direction. Specifically, they are chosen to satisfy
\begin{eqnarray}
    \psi^{(2)}_{sn}(t,{\bm x}) &\propto& \mathrm{e}^{-i\xi^2_2/2},\quad \xi_2\rightarrow-\infty,\nonumber\\
    \psi^{(4)}_{sn}(t,{\bm x}) &\propto& \mathrm{e}^{i\xi^2_2/2},\quad \xi_2\rightarrow-\infty,\nonumber\\
    \psi^{(3)}_{sn}(t,{\bm x}) &\propto& \mathrm{e}^{i\xi^2_2/2},\quad \xi_2\rightarrow+\infty,\nonumber\\
    \psi^{(1)}_{sn}(t,{\bm x}) &\propto& \mathrm{e}^{-i\xi^2_2/2},\quad \xi_2\rightarrow+\infty.
\end{eqnarray}

As shown in Appendix~\ref{app:mode-orthonormality}, the asymptotically defined stationary modes satisfy the generalized orthonormality relations
\begin{eqnarray}
    \left(\psi^{(i)}_{sn}(t,\bm{x}|p_0,p_y),\psi^{(i)}_{s'n'}(t,\bm{x}|p_0',p_y')\right) &=& \delta_{ss'}\delta_{nn'}\delta(p_0-p'_0)\delta(p_y-p'_y),\quad i=1,2,3,4\nonumber\\
    \left(\psi^{(1)}_{sn}(t,\bm{x}|p_0,p_y),\psi^{(4)}_{s'n'}(t,\bm{x}|p_0',p_y')\right) &=&\left(\psi^{(2)}_{sn}(t,\bm{x}|p_0,p_y),\psi^{(3)}_{s'n'}(t,\bm{x}|p_0',p_y')\right)=0,  
\label{ortho}
\end{eqnarray}
and the completeness relations
\begin{eqnarray}
&& \sum_{i\in\{1,4\}}\sum_{ns}\int dp_0dp_y \psi^{(i)}_{sn}(t,\bm{x}|p_0,p_y)\psi^{(i)\dagger}_{sn}(t,\bm{x}'|p_0,p_y)\nonumber\\
&=& \sum_{i\in\{2,3\}}\sum_{ns}\int dp_0dp_y \psi^{(i)}_{sn}(t,\bm{x}|p_0,p_y)\psi^{(i)\dagger}_{sn}(t,\bm{x}'|p_0,p_y)= \delta^{(3)}(\bm {x}-\bm{x}')I_4, 
\label{complete}
\end{eqnarray}
where the inner product is defined on the equal-time hypersurface
\begin{equation}
    \left(\psi,\phi\right) = \int d^3\bm{x}\psi^\dagger(t,\bm{x})\phi(t,\bm{x}).
\end{equation}
Thus, $\{\psi^{(1)},\psi^{(4)}\}$ and $\{\psi^{(3)},\psi^{(2)}\}$ constitute two orthonormal and complete mode bases, which are identified respectively as the in- and out-mode bases,
\begin{eqnarray}
    && u_{sn}^{\text{in}} = \psi_{sn}^{(1)},\qquad v_{sn}^{\text{in}}=\psi_{sn}^{(4)},\nonumber\\
    && u_{sn}^{\text{out}} = \psi_{sn}^{(3)},\qquad v_{sn}^{\text{out}} = \psi_{sn}^{(2)}.
\end{eqnarray}
The quantized Dirac field can therefore be expanded in either basis as
\begin{eqnarray}
\hat\psi(t,\bm{x}) &=& \sum_{sn} \int d p_0 d p_y \left[ u_{sn}^{\text{in}}(t,\bm{x}|p_0,p_y)\hat{a}^{\text{in}}_{sn}(p_0,p_y) +  v_{sn}^{\text{in}}(t,\bm{x}|p_0,p_y)\hat{b}^{\text{in}\dagger}_{sn}(p_0,p_y) \right]\nonumber\\
&=& \sum_{sn} \int d p_0 d p_y \left[u_{sn}^{\text{out}}(t,\bm{x}|p_0,p_y) \hat{a}^{\text{out}}_{sn}(p_0,p_y) +  v_{sn}^{\text{out}}(t,\bm{x}|p_0,p_y) \hat{b}^{\text{out}\dagger}_{sn}(p_0,p_y) \right],
\label{eq:field-in-out-expansion}
\end{eqnarray}
where the creation and annihilation operators satisfy
\begin{equation}
    \left\{\hat{a}^{\text{in}}_{sn}(p_0,p_y), \hat{a}^{\text{in}\dagger}_{s'n'}(p_0',p_y')\right\} = 
       \left\{\hat{b}^{\text{in}}_{sn}(p_0,p_y), \hat{b}^{\text{in}\dagger}_{s'n'}(p_0',p_y')\right\} = \delta_{nn'}\delta_{ss'}    \delta(p_0-p'_0)\delta(p_y-p'_y),
\end{equation}
and the other anti-commutation relations in the in-mode basis vanish. Analogous relations hold in the out-mode basis. 

The two normalized bases $\{u_{sn}^{\text{in}},v_{sn}^{\text{in}}\}$ and $\{u_{sn}^{\text{out}},v_{sn}^{\text{out}}\}$ are related to each other by a Bogoliubov transformation, 
\begin{eqnarray}
	&& u_{sn}^{\text{out}}(p_0,p_y) = \alpha_n u_{sn}^{\text{in}}(p_0,p_y) + \beta_n v_{sn}^{\text{in}}(p_0,p_y),\nonumber\\
	&& v_{sn}^{\text{out}}(p_0,p_y) = -\beta_n u_{sn}^{\text{in}}(p_0,p_y) + \alpha_n v_{sn}^{\text{in}}(p_0,p_y),
\label{eq:mode-bogoliubov-transformation}
\end{eqnarray}
the Bogoliubov coefficients are chosen to be real and non-negative,
\begin{equation}
    \alpha_n = \sqrt{1-\mathrm{e}^{-\pi\mu_n}}, \qquad \beta_n = \mathrm{e}^{-\pi\mu_n/2}
\label{eq:real-bogoliubov-coefficients}
\end{equation}
satisfying the relation
\begin{equation}
    \alpha_n^2+\beta_n^2=1.
\label{eq:fermionic-bogoliubov-unitarity}
\end{equation}
The corresponding creation and annihilation operators are then related by
\begin{eqnarray}
    && \hat a_{sn}^{\text{out}}(p_0,p_y) = \alpha_n\hat a_{sn}^{\text{in}}(p_0,p_y) + \beta_n\hat b_{sn}^{\text{in}\dagger}(p_0,p_y), \nonumber\\
    && \hat b_{sn}^{\text{out}\dagger}(p_0,p_y) = -\beta_n\hat a_{sn}^{\text{in}}(p_0,p_y) + \alpha_n\hat b_{sn}^{\text{in}\dagger}(p_0,p_y),
\label{eq:operator-bogoliubov-transformation}
\end{eqnarray}
and the in- and out-vacua are defined by
\begin{eqnarray}
    && \hat a_{sn}^{\text{in}}(p_0,p_y)\lvert 0,\text{in}\rangle = \hat b_{sn}^{\text{in}}(p_0,p_y)\lvert 0,\text{in}\rangle =0, \nonumber\\
    && \hat a_{sn}^{\text{out}}(p_0,p_y)\lvert 0,\text{out}\rangle = \hat b_{sn}^{\text{out}}(p_0,p_y)\lvert 0,\text{out}\rangle =0.
\label{eq:in-out-vacuum-definitions}
\end{eqnarray}

We now calculate the causal in--out propagator, which is defined by
\begin{equation}
    S(x,x')
    =-i\frac{\langle 0,\text{out}|\hat T\hat\psi(x)\hat{\bar\psi}(x')|0,\text{in}\rangle}{\langle 0,\text{out}|0,\text{in}\rangle}.
\label{eq:causal-propagator-definition}
\end{equation}
With the real phase convention of Eq.~\eqref{eq:real-bogoliubov-coefficients}, the non-vanishing mixed vacuum contractions are
\begin{eqnarray}
    \frac{\langle 0,\text{out}\lvert \hat a_{sn}^{\text{out}}(p_0,p_y)\hat a_{s'n'}^{\text{in}\dagger}(p_0',p_y')\rvert 0,\text{in}\rangle}{\langle 0,\text{out}\vert 0,\text{in}\rangle}
    &=& \frac{\langle 0,\text{out}\lvert \hat b_{sn}^{\text{out}}(p_0,p_y)\hat b_{s'n'}^{\text{in}\dagger}(p_0',p_y')\rvert 0,\text{in}\rangle}{\langle 0,\text{out}\vert 0,\text{in}\rangle}\nonumber\\
    &=& \frac{1}{\alpha_n}\delta_{ss'}\delta_{nn'}\delta(p_0-p'_0)\delta(p_y-p'_y).
\label{eq:mixed-in-out-contractions}
\end{eqnarray}
Substituting the field expansions in Eq.~\eqref{eq:field-in-out-expansion} into Eq.~\eqref{eq:causal-propagator-definition}, we obtain
\begin{eqnarray}
    S(x,x') &=& -i\sum_{sn}\int\frac{dp_0dp_y}{\sqrt{1-\mathrm{e}^{-\pi\mu_n}}}\Big[\theta(t-t')\psi_{sn}^{(3)}(t,\bm{x}|p_0,p_y)\bar\psi_{sn}^{(1)}(t',\bm{x}'|p_0,p_y)\nonumber\\
    && -\theta(t'-t)\psi_{sn}^{(4)}(t,\bm{x}|p_0,p_y)\bar\psi_{sn}^{(2)}(t',\bm{x}'|p_0,p_y)\Big]
    \label{eq:propagator}
\end{eqnarray}
Using the Bogoliubov transformations \eqref{eq:mode-bogoliubov-transformation}-\eqref{eq:operator-bogoliubov-transformation} together with the completeness relation~\eqref{complete}, one can verify that the propagator satisfies 
\begin{equation}
    \left(i\gamma^\mu D_\mu^x-m\right)S(x,x') = \delta^{(4)}(x-x')I_4.
\end{equation}

\subsection{Proper-time representation of the propagator}
\label{sec:proper-time}
We now convert the mode sum in Eq.~\eqref{eq:propagator} into a proper-time integral. This provides a direct connection between the stationary-mode construction and the Schwinger representation. Substituting the modes and collecting their normalization factors gives
\begin{equation}
    S(x,x')=-{i} \sum_{n, s} \int {d} p_{0} {d} p_{y} \mathcal{N}_{n} \mathrm{e}^{-{i} p_{0} (t-t')+{i} p_{y} (y-y')}\left[\theta(t-t') \mathcal{B}_{+}-\theta(t'-t) \mathcal{B}_{-}\right] 
    \label{eq:propagator1}
\end{equation}
with the factor
\begin{equation}
    \mathcal{N}_{n}=\sqrt{\frac{eB}{2 \pi}} \frac{1}{4 \pi eE\left(\mathrm{e}^{\pi \mu_{n}}-1\right)}
\end{equation}
and the kernels
\begin{eqnarray}
\mathcal{B}_\mp &=& 2 i eE/\Gamma\left(-i \mu_{n}/2\right) D_{\nu_{-}}(\mp\eta) D_{\nu_{+}}\left(\mp\eta^{\prime}\right) U_{sn}(x) \bar{U}_{sn}(x')\nonumber\\ 
&& \pm(1+i) \sqrt{eE}/\Gamma\left(-i \mu_{n}/2\right) D_{\nu_{-}}(\mp\eta) D_{\nu_{-}}\left(\mp\eta^{\prime}\right) U_{sn}(x) \bar{V}_{sn}(x')\nonumber \\
&& \pm(1+i) \sqrt{eE}/\Gamma\left(1-i \mu_{n}/2\right) D_{\nu_{+}}(\mp\eta) D_{\nu_{+}}\left(\mp\eta^{\prime}\right) V_{sn}(x) \bar{U}_{sn}(x')\nonumber\\ 
&& +1/\Gamma\left(1-i \mu_{n}/2\right) D_{\nu_{+}}(\mp\eta) D_{\nu_{-}}\left(\mp\eta^{\prime}\right) V_{sn}(x) \bar{V}_{sn}(x'),
\end{eqnarray}
where $\nu_\pm, \eta$ and $\eta'$ are defined as
\begin{eqnarray}
&& \nu_+=-i\mu_n/2,\quad \nu_-=-i\mu_n/2-1,\nonumber\\
&& \eta=(1-i)\sqrt{eE}(z-\frac{p_0}{eE}),\quad \eta^{\prime}=(1-i)\sqrt{eE}(z^{\prime}-\frac{p_0}{eE}).
\end{eqnarray}

To perform the spin sum in Eq.~\eqref{eq:propagator1}, it is convenient to introduce the projectors
\begin{equation}
    \mathcal{P}_{ \pm}=\frac{I_{4} \pm \gamma^{0} \gamma^{3}}{2}, \quad \Pi_{ \pm}=\frac{I_{4} \pm 2 \Sigma_{z}}{2},
\end{equation}
together with the transverse kernel $\mathcal{H}_{n}$ and differential operator $Q_{\perp}$
\begin{eqnarray}
&& \mathcal{H}_{n}\left(x, x^{\prime} ; p_{y}\right)  =\varphi_{n-1}\left(\xi_{1}\right) \varphi_{n-1}\left(\xi_{1}^{\prime}\right) \Pi_{+}+\varphi_{n}\left(\xi_{1}\right) \varphi_{n}\left(\xi_{1}^{\prime}\right) \Pi_{-},\nonumber \\
&& Q_{\perp} = i \gamma^{1} \partial_{x}-\gamma^{2}\left(p_{y}+eB x\right)
\end{eqnarray}
with $\xi_1'=\sqrt{eB}(x'+\frac{p_y}{eB})$. The transverse spinors in Eq.~\eqref{eq:propagator1} then satisfy
\begin{eqnarray}
&& \sum_{s= \pm} U_{s n}(x) \bar{U}_{s n}\left(x^{\prime}\right) = \mathcal{H}_{n} \mathcal{P}_{+} \gamma^{0},\nonumber \\
&& \sum_{s= \pm} U_{s n}(x) \bar{V}_{s n}\left(x^{\prime}\right) = \left(m+Q_{\perp}\right) \mathcal{H}_{n} \mathcal{P}_{+},\nonumber \\
&& \sum_{s= \pm} V_{s n}(x) \bar{U}_{s n}\left(x^{\prime}\right) = \left(m+Q_{\perp}\right) \mathcal{H}_{n} \mathcal{P}_{-},\nonumber \\
&& \sum_{s= \pm} V_{s n}(x) \bar{V}_{s n}\left(x^{\prime}\right) = M_n^2 \mathcal{H}_{n} \mathcal{P}_{-} \gamma^{0},
\end{eqnarray}
with the transverse energy $M_n=\sqrt{m^2+2neB}$. Combining these identities with the recurrence relations for the parabolic cylinder functions \eqref{recurrence relations of D}, we find that the propagator can be expressed as the action of a Dirac operator on a kernel $G(x,x')$,
 \begin{equation}
     S\left(x, x^{\prime }\right)=\left(i \gamma^{\mu} D_{\mu}^x+m\right) G\left(x, x^{\prime }\right)
     \label{eq:StoG}
 \end{equation}
with 
 \begin{equation}
G\left(x, x^{\prime }\right)= 2 \pi i(1+i) \sqrt{eE} \sum_{n}   \int {d} p_{y} \mathrm{e}^{i p_{y} (y-y')} \mathcal{N}_{n}\mathcal{H}_{n} 
\left[F_{\nu_{-}}/\Gamma\left(-i \mu_{n}/2\right) \mathcal{P}_{+}+F_{\nu_{+}}/\Gamma\left(1-i \mu_{n}/2\right) \mathcal{P}_{-}\right],
\label{eq:G}
 \end{equation}
where the function $F_{\nu}$ is defined by
\begin{equation}
F_{\nu}\left(t-t' ; z, z^{\prime}\right)=\int \frac{{d} p_{0}}{2 \pi} {\mathrm e}^{-i p_{0} (t-t')}  \left[\theta(t-t') D_{\nu}(\eta) D_{\nu}\left(\eta^{\prime}\right)
+\theta(t'-t) D_{\nu}(-\eta) D_{\nu}\left(-\eta^{\prime}\right)\right],
\end{equation}
which leads to 
\begin{eqnarray}
    F_{\nu_\mp}\left(t-t' ; z, z^{\prime}\right) &=&\mp {\mathrm{e}^{ \pi \mu_n/2} (eE)^{3 / 2}\over 2 \sqrt{-i \pi} \Gamma(1/2\pm 1/2+i\mu_n/2)} \mathrm{e}^{-ieE/2(z+z')(t-t')}\int_{0}^{+\infty} {d s\over \sinh (eE s)}\nonumber\\ 
    &&\times \mathrm{e}^{\left[{-i(M_n^2-i0^+)s\mp eEs-i eE/4 \operatorname{coth}(eE s)\left[(t-t')^{2}-(z-z')^{2}\right]}\right]}.
    \label{eq:F_nu_minus}
\end{eqnarray}
Substituting Eqs.~\eqref{eq:F_nu_minus} into
Eq.~\eqref{eq:G}, we obtain
 \begin{eqnarray}
G(x,x') &=&-\frac{i eE \sqrt{eB}}{8 \pi^{2}} \mathrm{e}^{-ieE/2(z+z')(t-t')} \sum_{n=0}^{\infty} \int {d} p_{y} \mathrm{e}^{i p_{y} (y-y')} \int_{0}^{+\infty} \frac{d s}{\sinh (eE s)}\nonumber\\
&&\times \mathrm{e}^{-i\left(M_{n}^{2}-i 0^+\right) s} 
 \mathrm{e}^{-i eE/4 \operatorname{coth}(eE s)\left[(t-t')^{2}-(z-z')^{2}\right]} \mathcal{H}_{n} \mathrm{e}^{-eE s \gamma^{0} \gamma^{3}} .
 \label{G2}
 \end{eqnarray}
At this stage, only the Landau-level sum and the transverse-momentum integral in Eq.~\eqref{G2} remain to be evaluated. The Landau-level sum can be written as
 \begin{eqnarray}
\sum_{n=0}^{\infty} \mathrm{e}^{-2 i n eB s} \mathcal{H}_{n} &=& \sum_{\ell=0}^{\infty} \mathrm{e}^{-2 i \ell eB s} \varphi_{\ell}\left(\xi_{1}\right) \varphi_{\ell}\left(\xi_{1}^{\prime}\right)\left(\mathrm{e}^{-2 i eB s} \Pi_{+}+\Pi_{-}\right)\nonumber \\
&=& \sum_{\ell=0}^{\infty} \mathrm{e}^{-i(2 \ell+1) eB s} \varphi_{\ell}\left(\xi_{1}\right) \varphi_{\ell}\left(\xi_{1}^{\prime}\right) \mathrm{e}^{-2 i eB s \Sigma_{z}} .
\end{eqnarray}
Setting $r=\mathrm{e}^{-2 i eB s}$ and $s\to s-i0^+$, we have $|r|<1$, and applying Mehler's formula
 \begin{equation}
     \sum_{\ell=0}^{\infty} r^{\ell} \varphi_{\ell}\left(\xi_{1}\right) \varphi_{\ell}\left(\xi_{1}^{\prime}\right)=\frac{1}{\sqrt{\pi\left(1-r^{2}\right)}} \mathrm{e}^{ -\left[\left(1+r^{2}\right)\left(\xi_{1}^{2}+\xi_{1}^{\prime 2}\right)-4 r \xi_{1} \xi_{1}^{\prime}\right]/\left[2\left(1-r^{2}\right)\right]},
 \end{equation}
we obtain
 \begin{eqnarray}
&& \sqrt{eB} \int \frac{{d} p_{y}}{2 \pi} \mathrm{e}^{i p_{y} (y-y')} \sum_{\ell=0}^{\infty} \mathrm{e}^{-i(2 \ell+1) eB (s-i0^+)} \varphi_{\ell}\left(\xi_{1}\right) \varphi_{\ell}\left(\xi_{1}^{\prime}\right) \nonumber\\
&=& \frac{eB}{4 \pi i \sin \left(eB \left(s-i0^+\right)\right)} \mathrm{e}^{-i eB/2(x+x^{\prime})(y-y')+i eB/4 \cot \left(eB \left(s-i0^+\right)\right)\left[(x-x')^{2}+(y-y')^{2}\right]} .
 \end{eqnarray}
Combining this result with Eq.~\eqref{G2} gives
 \begin{eqnarray}
G\left(x, x^{\prime }\right) &=& -\frac{1}{(4 \pi)^{2}}\mathrm{e}^{-ieE/2(z+z')(t-t')-i eB/2(x+x^{\prime})(y-y')} \int^{+\infty}_{0} {d} s \frac{eE }{\sinh (eE s)} \frac{eB }{\sin (eB (s-i0^+))} \mathrm{e}^{-i\left(m^{2}-i 0^+\right) s}\nonumber \\
&& \times \mathrm{e}^{-i eE/4 \operatorname{coth}(eE s)\left[(t-t')^{2}-(z-z')^{2}\right]+i eB/4 \cot \left(eB \left(s-i0^+\right)\right)\left[(x-x')^{2}+(y-y')^{2}\right]-eE s \gamma^{0} \gamma^{3}-2 i eB s \Sigma_{z}} .
\end{eqnarray}
Using Eq.~\eqref{eq:StoG}, we ultimately arrive at an expression for the propagator in terms of a proper-time integral, which is identical to that obtained from the Schwinger proper-time method~\cite{Schwinger:1951nm}:
 \begin{eqnarray}
S\left(x, x^{\prime }\right) &=& -\frac{1}{(4 \pi)^{2}} \mathrm{e}^{-ieE/2(z+z')(t-t')-i eB/2(x+x^{\prime})(y-y')}\int^{+\infty}_{0} d s \frac{eE }{\sinh (eEs )} \frac{eB }{\sin (eB (s-i0^+))}\nonumber\\
&& \times \mathrm{e}^{-i\left(m^{2}-i 0^+\right) s-eE s \gamma^{0} \gamma^{3}-2 i eB s \Sigma_{z}}\nonumber \\
&& \times \mathrm{e}^{-i eE/4 \operatorname{coth}(eE s)\left[(t-t')^{2}-(z-z')^{2}\right]+i eB/4 \cot \left(eB \left(s-i0^+\right)\right)\left[(x-x')^{2}+(y-y')^{2}\right]}\nonumber  \\
&& \times\left[m+\frac{eE \mathrm{e}^{eE s \gamma^{0} \gamma^{3}}}{2 \sinh (eE s)}\left[\gamma^{0} (t-t')-\gamma^{3} (z-z')\right]-\frac{eB \mathrm{e}^{2 i eB s \Sigma_{z}}}{2 \sin (eB (s-i0^+))}\left[\gamma^{1} (x-x')+\gamma^{2} (y-y')\right]\right] .
 \end{eqnarray}
 
\section{Perpendicular electric and magnetic fields}
\label{sec:perpendicular-fields}
The exact solutions to the Dirac equation in perpendicular electric and magnetic fields were previously investigated in Ref.~\cite{Canuto:1969ic}, where the authors focused primarily on the magnetically dominated regime ($B > E$). In this section, we focus on the electrically dominated case ($E > B$).

We consider a uniform magnetic field oriented along the $-x$ direction and a uniform electric field along the $-z$ direction. The corresponding four-potential in the laboratory frame $K$ can be chosen as $A_\mu=(Ez,0,-Bz,0)$. The covariance of the Dirac equation under a Lorentz transformation $x'^\mu = \Lambda^\mu_{\phantom{\mu}\nu}x^\nu$ ensures that
\begin{equation}
	\left(i\gamma^{\mu}D_{\mu}-m\right)\psi(x) = 0 \iff \left(i\gamma^{\mu}D'_{\mu}-m\right)\psi'(x') = 0
	\label{Dirac perpendicular}
\end{equation}
with $D'_{\mu}=\partial'_{\mu}+ieA'_{\mu}$. The wave functions in the two frames are related to each other via the inverse spinor transformation.

We now take a Lorentz boost along the $y$ axis with velocity $\beta=B/E$, the associated Lorentz transformation matrix is
\begin{equation}
	\Lambda^\mu_{\phantom{\mu}\nu} = \begin{pmatrix} 
		\gamma & 0 & -\gamma\beta & 0 \\ 
		0 & 1 & 0 & 0 \\ 
		-\gamma\beta & 0 & \gamma & 0 \\ 
		0 & 0 & 0 & 1 
	\end{pmatrix},
\end{equation}
where the Lorentz factor is 
\begin{equation}
\gamma=1/\sqrt{1-\beta^2}=E/\sqrt{E^2-B^2}.
\end{equation}
In the case of a magnetic field weaker than the electric field, $E>B$, the equation for the stationary state in the boosted frame $K'$ is reduced to the equation in a pure electric field $E'=\sqrt{E^{2}-B^{2}}$, and the four-potential becomes $A'_\mu=(E'z',0,0,0)$. In this case, the spinor transformation operator is
\begin{equation}
S(\Lambda)=\mathrm{e}^{-{\eta\over 2}\gamma^0\gamma^2}
\end{equation}
with the rapidity $\eta=\operatorname{artanh}(\beta)$, and the four linearly independent solutions of the Dirac equation~\eqref{Dirac perpendicular} in the $K'$ frame are known, see Ref.~\cite{Zhao:2025prd},
\begin{eqnarray}
	{\psi'}_1^{\pm}(t',\bm{x}') &=&\frac{\mathrm{e}^{-i(p_0' t'-p_x'x'-p_y'y')}}{\sqrt{2}}
	\begin{pmatrix}
		\pm(i-1)\sqrt{eE'}D_{i\lambda'/2}(\pm\zeta)+mD_{i\lambda'/2-1}(\pm\zeta) \\
		{p}_+^\prime D_{i\lambda'/2-1}(\pm\zeta) \\
		\pm(i-1)\sqrt{eE'}D_{i\lambda'/2}(\pm\zeta)-mD_{i\lambda'/2-1}(\pm\zeta) \\
		p_+^\prime D_{i\lambda'/2-1}(\pm\zeta) \\
	\end{pmatrix},\nonumber\\
	{\psi'}_2^{\pm}(t',\bm{x}') &=&\frac{\mathrm{e}^{-i(p_0' t'-p_x'x'-p_y'y')}}{\sqrt{2}}
	\begin{pmatrix}
		-p_-^\prime D_{i\lambda'/2-1}(\pm\zeta) \\
		\pm (i-1)\sqrt{eE'}D_{i\lambda'/2}(\pm\zeta)+mD_{i\lambda'/2-1}(\pm\zeta) \\
		p_-^\prime D_{i\lambda'/2-1}(\pm\zeta) \\
		\pm(1-i)\sqrt{eE'}D_{i\lambda'/2}(\pm\zeta)+mD_{i\lambda'/2-1}(\pm\zeta) \\
	\end{pmatrix}
\end{eqnarray}
with the conserved energy and momenta $p'=(p_{0}', p_x', p_y')$ and the parameters and variables, 
\begin{eqnarray}
&& \lambda' = \frac{{p_x'}^2+{p_y'}^2+m^2}{eE'},\nonumber \\
&& \zeta = (1+i)\sqrt{eE'}\left(z'-\frac{p_0'}{eE'}\right),\nonumber\\
&& p_\pm^\prime = p_x^\prime\pm i p_y^\prime.
\label{conserved quantum numbers}
\end{eqnarray}

To obtain the wave functions $\psi(t,\bm{x})$ in the laboratory frame $K$, we perform the inverse Lorentz transformation $\psi(t,\bm{x}) = S^{-1}(\Lambda)\psi'(t',\bm{x}')$. Since $p_0$, $p_x$ and $p_y$ are conserved, the kinematic variables in the $K'$ frame are related to them via the following relations,
\begin{eqnarray}
&& p_0' = \gamma(p_0 - \beta p_y),\nonumber  \\
&& p_x' = p_x,\nonumber  \\
&& p_y' = \gamma(p_y - \beta p_0),
\label{eq:momentum_transform}
\end{eqnarray}
and the longitudinal coordinate remains invariant, $z'=z$. By substituting Eq.~\eqref{eq:momentum_transform} into Eq.~\eqref{conserved quantum numbers}, we have
\begin{eqnarray}
&& \lambda' = \frac{p_x^2 + \gamma^2(p_y - \beta p_0)^2 + m^2}{eE'}, \nonumber \\
&& \zeta = (1+i)\sqrt{eE'}\left(z - \frac{\gamma(p_0 - \beta p_y)}{eE'}\right), \nonumber \\
&& p_\pm^\prime = p_x \pm i \gamma(p_y - \beta p_0).
\end{eqnarray}

The inverse spinor representation of the Lorentz boost is given by
\begin{equation}
	S^{-1}(\Lambda) = \mathrm{e}^{{\eta\over 2}\gamma^0\gamma^2} = \cosh\left(\frac{\eta}{2}\right) I + \sinh\left(\frac{\eta}{2}\right)\gamma^0\gamma^2,
\end{equation}
in the standard Dirac representation, it takes the explicit matrix form
\begin{equation}
	S^{-1}(\Lambda) = \begin{pmatrix}
		\cosh\frac{\eta}{2} & 0 & 0 & -i\sinh\frac{\eta}{2} \\
		0 & \cosh\frac{\eta}{2} & i\sinh\frac{\eta}{2} & 0 \\
		0 & -i\sinh\frac{\eta}{2} & \cosh\frac{\eta}{2} & 0 \\
		i\sinh\frac{\eta}{2} & 0 & 0 & \cosh\frac{\eta}{2}
	\end{pmatrix}.
\end{equation}
Operating with $S^{-1}(\Lambda)$ on the $K'$-frame solution $\psi'(t',{\bm x}')$, we finally obtain the wave functions $\psi(t,{\bm x})$ of the Dirac equation in the laboratory frame,
\begin{eqnarray}
	\psi_1^{\pm}(t,\bm{x}) &=& \frac{\mathrm{e}^{-i(p_0 t-p_x x-p_y y)}}{\sqrt{2}}
	\begin{pmatrix}
		\pm(i-1)\sqrt{eE'}\cosh\frac{\eta}{2} D_{i\lambda'/2}(\pm\zeta) + \left(m\cosh\frac{\eta}{2} - i p_+^\prime\sinh\frac{\eta}{2}\right) D_{i\lambda'/2-1}(\pm\zeta) \\
		\pm(i-1)\sqrt{eE'}i\sinh\frac{\eta}{2} D_{i\lambda'/2}(\pm\zeta) + \left(-im\sinh\frac{\eta}{2} + p_+^\prime\cosh\frac{\eta}{2}\right) D_{i\lambda'/2-1}(\pm\zeta) \\
		\pm(i-1)\sqrt{eE'}\cosh\frac{\eta}{2} D_{i\lambda'/2}(\pm\zeta) - \left(m\cosh\frac{\eta}{2} + i p_+^\prime\sinh\frac{\eta}{2}\right) D_{i\lambda'/2-1}(\pm\zeta) \\
		\pm(i-1)\sqrt{eE'}i\sinh\frac{\eta}{2} D_{i\lambda'/2}(\pm\zeta) + \left(im\sinh\frac{\eta}{2} + p_+^\prime\cosh\frac{\eta}{2}\right) D_{i\lambda'/2-1}(\pm\zeta)
	\end{pmatrix},\nonumber\\
	\psi_2^{\pm}(t,\bm{x}) &=& \frac{\mathrm{e}^{-i(p_0 t-p_x x-p_y y)}}{\sqrt{2}}
	\begin{pmatrix}
		\pm(i-1)\sqrt{eE'}i\sinh\frac{\eta}{2} D_{i\lambda'/2}(\pm\zeta) - \left(im\sinh\frac{\eta}{2} + p_-^\prime\cosh\frac{\eta}{2}\right) D_{i\lambda'/2-1}(\pm\zeta) \\
		\pm(i-1)\sqrt{eE'}\cosh\frac{\eta}{2} D_{i\lambda'/2}(\pm\zeta) + \left(m\cosh\frac{\eta}{2} + i p_-^\prime\sinh\frac{\eta}{2}\right) D_{i\lambda'/2-1}(\pm\zeta) \\
		\mp(i-1)\sqrt{eE'}i\sinh\frac{\eta}{2} D_{i\lambda'/2}(\pm\zeta) - \left(im\sinh\frac{\eta}{2} - p_-^\prime\cosh\frac{\eta}{2}\right) D_{i\lambda'/2-1}(\pm\zeta) \\
		\mp(i-1)\sqrt{eE'}\cosh\frac{\eta}{2} D_{i\lambda'/2}(\pm\zeta) + \left(m\cosh\frac{\eta}{2} - i p_-^\prime\sinh\frac{\eta}{2}\right) D_{i\lambda'/2-1}(\pm\zeta)
	\end{pmatrix}.
	\label{eq:psi2_lab}
\end{eqnarray}

\section{Arbitrary electromagnetic fields}
\label{sec:arbitrary-fields}
We now extend the treatment of perpendicular fields to arbitrary electromagnetic fields. To be specific, we take the direction of the electric field as the $-z$ axis and the direction of the transverse component of the magnetic field as the $-x$ axis. Under this choice we have the fields
\begin{equation}
\bm {E} = (0, 0, -E),\qquad \bm {B} = (-B_x, 0, -B_z)	
\end{equation} 
and the potential
\begin{equation}
	A_{\mu} = (Ez, 0, -B_{x}z + B_{z}x, 0)
\end{equation}
in the laboratory frame $K$. To find the analytic solution to the Dirac equation, we implement a two-step spacetime transformation: a Lorentz boost to render the fields parallel, followed by a spatial rotation to align them with the $-z$ axis.

We first apply a Lorentz boost along the $y$ axis with velocity $\beta$ to an intermediate frame $K'$. The transformation matrix $\Lambda^\mu_{\phantom{\mu}\nu}$ takes the same form as in the perpendicular case. Under this boost, the transformed electric and magnetic fields in the $K'$ frame possess the components
\begin{eqnarray}
&&	E'_x = -\gamma \beta B_z, \quad E'_y = 0, \quad E'_z = -\gamma(E - \beta B_x), \nonumber \\
&&	B'_x = -\gamma(B_x - \beta E), \quad B'_y = 0, \quad B'_z = -\gamma B_z
\end{eqnarray}
with $\gamma = 1/\sqrt{1-\beta^2}$. To render the fields mutually parallel in the $K'$ frame, we require 
\begin{equation}
\bm{E}' \times \bm{B}' = 0,
\end{equation}
which yields the condition
\begin{equation}
	E'_z B'_x - E'_x B'_z = 0.
\end{equation}
This leads to a quadratic equation for the boost velocity, 
\begin{equation}
	E B_x \beta^2 - (E^2 + B^2)\beta + E B_x = 0
\end{equation}
with $B^2 = B_x^2 + B_z^2$. To ensure a physical velocity $|\beta| < 1$, the required boost velocity is uniquely determined as
\begin{equation}
	\beta = \frac{E^2 + B^2 - \sqrt{(E^2 + B^2)^2 - 4E^2 B_{x}^2}}{2E B_{x}}.
\end{equation}

The fields $\bm{E}'$ and $\bm{B}'$ are now parallel and lie in the $x'-z'$ plane. We then perform a spatial rotation around the $y'$ axis by an angle $\theta$ to arrive at the final frame $K''$, where both fields are aligned along the $-z''$ direction. The transformation of contravariant four-vectors from $K'$ to $K''$, denoted by $A''^{\mu} = {{R}^{\mu}}_{\nu} A'^{\nu}$, is governed by the explicit $4 \times 4$ rotation matrix
\begin{equation}
    {{R}^{\mu}}_{\nu} = \begin{pmatrix} 
        1 & 0 & 0 & 0 \\ 
        0 & \cos\theta & 0 & -\sin\theta \\ 
        0 & 0 & 1 & 0 \\ 
        0 & \sin\theta & 0 & \cos\theta 
    \end{pmatrix}.
\label{eq:rotation_matrix}
\end{equation}
The rotation angle $\theta$ is determined by 
\begin{eqnarray}
&&    \sin\theta = -\frac{E'_x}{\sqrt{E^{'2}_x + E^{'2}_z}} = -\frac{B'_x}{\sqrt{B^{'2}_x + B^{'2}_z}},\nonumber\\ 
&&    \cos\theta = -\frac{E'_z}{\sqrt{E^{'2}_x + E^{'2}_z}} = -\frac{B'_z}{\sqrt{B^{'2}_x + B^{'2}_z}}.
\end{eqnarray}

In the $K''$ frame, the four-potential reads
\begin{equation}
A''_\mu = \left(\sqrt{E^{'2}_x + E^{'2}_z} z'', 0, \sqrt{B^{'2}_x + B^{'2}_z} x'', 0\right).
\end{equation}
Consequently, the Dirac equation in the frame $K''$ becomes structurally identical to the case of parallel fields investigated in Sec.~\ref{subsec:Solution-in-parallel}. The stationary solution can be directly expressed as
\begin{equation}
	\psi''_{sn}(t'',\bm{x}'') = \mathrm{e}^{-i(p''_0 t'' - p''_y y'')}\phi_{sn}(x'', z''),
\end{equation}
where the spinor components $\phi_{sn}(x'', z'')$ are given by Eqs.~\eqref{eq:parallel} under the substitutions $E \to \sqrt{E^{'2}_x + E^{'2}_z}$ and $B \to \sqrt{B^{'2}_x + B^{'2}_z}$.

Finally, the wave functions $\psi_{sn}(t,\bm{x})$ in the laboratory frame $K$ are reconstructed via the inverse spinor transformation,
\begin{equation}
	\psi_{sn}(t,\bm{x}) = S^{-1}(\Lambda) S^{-1}(R) \psi''_{sn}(t'', \bm{x}''),
\end{equation}
where the inverse boost operator $S^{-1}(\Lambda)$ and the inverse rotation operator $S^{-1}(R)$ are given explicitly in the standard Dirac representation by
\begin{eqnarray}
	S^{-1}(\Lambda) &=& \cosh\left(\frac{\eta}{2}\right) I_4 + \sinh\left(\frac{\eta}{2}\right)\gamma^0\gamma^2, \nonumber \\
	S^{-1}(R) &=& \cos\left(\frac{\theta}{2}\right) I_4 - \sin\left(\frac{\theta}{2}\right)\gamma^1\gamma^3
\end{eqnarray}
with the rapidity defined as $\eta = \operatorname{artanh}(\beta)$. 

To explicitly evaluate the wave function in the laboratory frame $K$, the transformed spacetime coordinates $(t'', \bm{x}'')$ and the conserved energy-momentum components $(p''_0, p''_y)$ appearing in $\psi''_{sn}(t'', \bm{x}'')$ should be expressed in terms of the laboratory variables, which are given by
\begin{eqnarray}
	x'' &=& x \cos\theta - z \sin\theta, \nonumber \\
	y'' &=& \gamma(y - \beta t), \nonumber \\
	z'' &=& x \sin\theta + z \cos\theta, \nonumber \\
	t'' &=& \gamma(t - \beta y),
\end{eqnarray}
and
\begin{eqnarray}
	p''_0 &=& \gamma(p_0 - \beta p_y), \nonumber \\
	p''_y &=& \gamma(p_y - \beta p_0).
\end{eqnarray}

\section{Conclusions}
\label{sec:conclusions}
In this work, we have constructed analytic solutions of the Dirac equation in several configurations of external electric and magnetic fields. For mutually parallel fields, the Dirac equation was separated into a transverse harmonic-oscillator equation and a longitudinal Weber equation. The resulting stationary spinors were used to construct two orthonormal and complete sets of in- and out-modes. We also obtained their Bogoliubov transformation and derived the causal in--out propagator. Its consistency with the Green function equation follows from the completeness relation of the stationary modes.

For perpendicular fields in the electrically dominated regime, $E>B$, the analytic solutions were obtained by transforming the pure-electric-field solutions back to the laboratory frame. We further extended this procedure to arbitrary electric and magnetic fields with a general relative orientation. A Lorentz boost was used to render the transformed fields parallel, followed by a spatial rotation that aligned them in a common direction. The laboratory-frame solutions were then reconstructed through the spinor transformations.

These analytical solutions and the causal in--out propagator provide a useful basis for investigating vacuum pair production and other strong-field quantum processes in general electromagnetic backgrounds.

\vspace{1cm}
\noindent {\bf Acknowledgment}:
This work is supported by Yantai University under Grant No.~2226001 and by the National Natural Science Foundation of China (NSFC) under Grant Nos.~12575149 and 12175165.

\appendix
\section{Properties of the parabolic cylinder functions}
\label{app:cylinder}
In this Appendix, we briefly review the definition and key properties of the parabolic cylinder function $D_\nu(z)$ used in the main text. It is a standard solution to the second-order linear ordinary differential equation, frequently referred to as Weber's equation~\cite{gradshteyn2014table},
\begin{equation}
    \frac{d^2 y}{dz^2} + \left(\nu + \frac{1}{2} - \frac{z^2}{4}\right)y = 0.
\end{equation}

For a given parameter $\nu$, the two linearly independent solutions can be chosen as $D_\nu(z)$ and $D_\nu(-z)$, provided $\nu$ is not a non-negative integer. Their linear independence is verified by the Wronskian~\cite{bateman1953higher},
\begin{equation}
    D_\nu(z)\frac{d}{dz}D_\nu(-z)-D_\nu(-z)\frac{d}{dz}D_\nu(z)=\frac{\sqrt{2\pi}}{\Gamma{(-\nu)}}.
\end{equation}
The function $D_\nu(z)$ satisfies the recurrence relations,
\begin{eqnarray}
&& D_{\nu+1}(z)-zD_\nu(z)+\nu D_{\nu-1}(z)=0,\nonumber\\
&& \frac{d}{dz}D_\nu(z)+\frac{1}{2}zD_\nu(z)-\nu D_{\nu-1}(z)=0,\nonumber\\
&& \frac{d}{dz}D_\nu(z)-\frac{1}{2}zD_\nu(z)+D_{\nu+1}(z)=0
\label{recurrence relations of D}
\end{eqnarray}
and the connection formulas,
\begin{eqnarray}
	D_\nu(z) &=& \frac{\Gamma(\nu+1)}{\sqrt{2\pi}} \left[\mathrm{e}^{\pi \nu i/2} D_{-\nu-1}(iz) + \mathrm{e}^{-\pi \nu i/2} D_{-\nu-1}(-iz) \right]\nonumber\\
	&=& \mathrm{e}^{-\nu\pi i} D_\nu(-z) + \frac{\sqrt{2\pi}}{\Gamma(-\nu)} \mathrm{e}^{-\pi(\nu+1)i/2} D_{-\nu-1}(iz)\nonumber \\
	&=& \mathrm{e}^{\nu\pi i} D_\nu(-z) + \frac{\sqrt{2\pi}}{\Gamma(-\nu)} \mathrm{e}^{\pi(\nu+1)i/2} D_{-\nu-1}(-iz),
\end{eqnarray}
and its asymptotic expansion for large arguments $|z|\gg 1$ and $|z|\gg |\nu|$ depends on the region of $z$,
\begin{eqnarray}
&&    D_\nu(z)\sim \mathrm{e}^{-z^2/4}z^\nu\ \ \ \text{for}\ \ \ |\arg(z)| < \frac{3}{4}\pi,\nonumber\\
&&    D_\nu(z)\sim \mathrm{e}^{-z^2/4}z^\nu-\frac{\sqrt{2\pi}}{\Gamma(-\nu)}\mathrm{e}^{i\pi\nu}\mathrm{e}^{z^2/4}z^{-\nu-1}\ \ \ \text{for}\ \ \ \frac{1}{4}\pi<\arg(z)<\frac{5}{4}\pi,\nonumber\\
&&    D_\nu(z)\sim \mathrm{e}^{-z^2/4}z^\nu-\frac{\sqrt{2\pi}}{\Gamma(-\nu)}\mathrm{e}^{-i\pi\nu}\mathrm{e}^{z^2/4}z^{-\nu-1}\ \ \ \text{for}\ \ \ -\frac{1}{4}\pi>\arg(z)>-\frac{5}{4}\pi.
\end{eqnarray}

\section{Properties of the harmonic oscillator functions}
\label{app:harmonic-oscillator}

In this Appendix, we summarize the properties of the normalized harmonic oscillator function $\varphi_n(\xi)$ used in the main text. It is defined as 
\begin{equation}
    \varphi_n(\xi) = \frac{1}{\pi^{1/4}\sqrt{2^n n!}} \mathrm{e}^{-\xi^2/2}H_n(\xi), \qquad n=0,1,2,\ldots 
\label{eq:harmonic-oscillator-function}
\end{equation}
with $H_n(\xi)$ being the Hermite polynomial. It obeys the harmonic oscillator equation
\begin{equation}
    \left[\frac{d^2}{d\xi^2}-\xi^2+2n+1\right]\varphi_n(\xi) = 0,
\end{equation}
and satisfies the recurrence relations
\begin{eqnarray}
    && \xi\varphi_n(\xi) = \sqrt{\frac{n}{2}}\varphi_{n-1}(\xi)+\sqrt{\frac{n+1}{2}}\varphi_{n+1}(\xi),\nonumber\\
    && \frac{d}{d\xi}\varphi_n(\xi) = \sqrt{\frac{n}{2}}\varphi_{n-1}(\xi)-\sqrt{\frac{n+1}{2}}\varphi_{n+1}(\xi)
\label{eq:harmonic-oscillator}
\end{eqnarray}
and the orthonormality and completeness relations,
\begin{eqnarray}
&& \int_{-\infty}^{+\infty}d\xi\varphi_n^*(\xi)\varphi_{n'}(\xi) = \delta_{nn'},\nonumber\\
&& \sum_{n=0}^{\infty}\varphi_n(\xi)\varphi_n^*(\xi') = \delta(\xi-\xi').
\label{eq:harmonic-oscillator-completeness}
\end{eqnarray}

\section{Magnetic polarization eigenstates}
\label{app:magnetic-polarization}
The label $s=\pm$ in the stationary modes~\eqref{asymptotically defined stationary modes} does not directly specify their conserved magnetic polarization. For parallel electric and magnetic fields, a suitable polarization basis can be constructed using the Sokolov--Ternov magnetic moment operator~\cite{sokolov1968synchrotron,baring2005spin},
\begin{equation}
    \hat{\mu}_{z}=m \mathrm{~S}_{z}+\rho_{y}\left(\mathrm{~S}_{x} \Pi_{y}-\mathrm{S}_{y} \Pi_{x}\right),
\end{equation}
 with 
\begin{equation}
    \mathrm{S}_{j}=\left(\begin{array}{cc}
\sigma_{j} & 0 \\
0 & \sigma_{j}
\end{array}\right), \quad \rho_{y}=-i\gamma^0\gamma^5=\left(\begin{array}{cc}
0 & -i I_{2} \\
i I_{2} & 0
\end{array}\right),\quad\Pi_{x}=-i \partial_{x}, \quad \Pi_{y}=-i \partial_{y}+e B x .
\end{equation}
This Hermitian operator commutes with the Hamiltonian in
Eq.~\eqref{eq:hamiltonian},
\begin{equation}
    [\hat H,\hat{\mu}_z]=0,
\end{equation}
and therefore defines a conserved magnetic polarization. By taking appropriate linear combinations of the wave functions in Eq.~\eqref{asymptotically defined stationary modes}, one can construct the corresponding eigenstates of $\hat{\mu}_z$, namely,
\begin{equation}
    \begin{aligned}
        \Psi_{+n}^{(i)}(t,\bm{x}|p_0,p_y)
        &=c_n\psi_{+n}^{(i)}(t,\bm{x}|p_0,p_y)-id_n\psi_{-n}^{(i)}(t,\bm{x}|p_0,p_y),\\
        \Psi_{-n}^{(i)}(t,\bm{x}|p_0,p_y)
        &=-id_n\psi_{+n}^{(i)}(t,\bm{x}|p_0,p_y)+c_n\psi_{-n}^{(i)}(t,\bm{x}|p_0,p_y),
    \end{aligned}
    \qquad n\geq1,
    \label{eq:magnetic-polarization-eigenstates}
\end{equation}
with
\begin{equation}
    c_n=\sqrt{\frac{M_n+m}{2M_n}},
    \qquad
    d_n=\sqrt{\frac{M_n-m}{2M_n}}.
\end{equation}
They satisfy
\begin{equation}
    \hat{\mu}_z\Psi_{\pm n}^{(i)}
    =\pm M_n\Psi_{\pm n}^{(i)}.
\end{equation}
At the lowest Landau level, $\varphi_{-1}=0$ implies $\psi_{+0}^{(i)}=0$. Only $\Psi_{-0}^{(i)}=\psi_{-0}^{(i)}$ remains, with eigenvalue $-m$.

\section{Normalization, orthogonality, and completeness of the stationary modes}
\label{app:mode-orthonormality}
In this Appendix, we establish the orthonormality and completeness of the stationary modes introduced in Eq.~\eqref{asymptotically defined stationary modes}. Since $p_0$ and $p_y$ are continuous quantum numbers, the stationary modes are generalized eigenfunctions rather than square-integrable states. All equalities involving Dirac delta functions are therefore understood in the sense of tempered distributions, or equivalently as zero-width limits of Gaussian wave packets.

\subsection{Orthonormality}
The transverse spinors $U$ and $V$ defined in Eq.~\eqref{eq:UandV} satisfy the orthonormality conditions,
\begin{eqnarray}
    && \int_{-\infty}^{+\infty}dx U^\dagger_{sn}(x)U_{s'n'}(x) = \frac{1}{\sqrt{eB}}\delta_{nn'}\delta_{ss'},\nonumber\\
    && \int_{-\infty}^{+\infty}dx V^\dagger_{sn}(x)V_{s'n'}(x) = \frac{M_n^2}{\sqrt{eB}}\delta_{nn'}\delta_{ss'},\nonumber\\
    && \int_{-\infty}^{+\infty}dx U^\dagger_{sn}(x)V_{s'n'}(x) = 0.
\label{eq:UV_orth_packet}
\end{eqnarray}

We introduce the functions
\begin{equation}
    F_1(z|p_0) = \begin{pmatrix} A_1(z|p_0)\\ M_n B_1(z|p_0) \end{pmatrix},
    \qquad F_4(z|p_0) = \begin{pmatrix} A_4(z|p_0)\\ M_n B_4(z|p_0) \end{pmatrix}
\label{eq:longitudinal_F_definition}
\end{equation}
which obey the equations, 
\begin{equation}
    \left[e E zI_2 +M_n\sigma_x -i\sigma_z\partial_z \right]F_i(z|p_0) = p_0F_i(z|p_0), \qquad i=1,4
\label{eq:longitudinal_first_order_packet}
\end{equation}
and introduce the finite-width packets,
\begin{equation}
    \Phi_i(t,z|\bar p_0,\sigma) = \int_{-\infty}^{+\infty}dp_0 G(p_0|\bar p_0,\sigma)\mathrm{e}^{-ip_0t}F_i(z|p_0),
\label{eq:time_dependent_longitudinal_packet}
\end{equation}
where $G(p_0|\bar p_0,\sigma)$ is the normalized Gaussian wave packet centered at $\bar p_0$ with width $\sigma>0$,
\begin{equation}
    G(p_0|\bar p_0,\sigma) = \frac{1}{\sqrt{2\pi}\sigma} \mathrm{e}^{-\frac{(p_0-\bar p_0)^2}{2\sigma^2}}
\label{eq:Gaussian_energy_kernel}
\end{equation}
satisfying the conditions, 
\begin{eqnarray}
    && \int_{-\infty}^{+\infty}dp_0 G(p_0|\bar p_0,\sigma) = 1,\nonumber\\
    && \lim_{\sigma\to 0^+}G(p_0|\bar p_0,\sigma) = \delta(p_0-\bar p_0).
\label{eq:Gaussian_energy_weak_limit}
\end{eqnarray}

Since the $p_0$-dependence of the functions $F_i(z|p_0)$ comes from the variable $\xi_2=\sqrt{e E}(z-p_0/(eE))$, we have the relation
\begin{equation}
    F_i(z|p_0) = F_i(z-p_0/(eE)|0).
\label{eq:stationary_translation_property}
\end{equation}
Setting $p_0=q+\bar p_0$ leads to
\begin{eqnarray}
    \Phi_i(t,z|\bar p_0,\sigma) &=& \mathrm{e}^{-i\bar p_0 t}\int_{-\infty}^{+\infty}dq G(q|0,\sigma)\mathrm{e}^{-iqt} F_i(z-(\bar p_0+q)/(eE)|0)\nonumber\\
    &=& \mathrm{e}^{-i\bar p_0 t}\Phi_i(t,z-\bar p_0/(e E)|0,\sigma),
\label{eq:translation_property_packet2}
\end{eqnarray}
and 
\begin{eqnarray}
    \left[e E zI_2+M_n\sigma_x-i\sigma_z\partial_z\right]\Phi_i(t,z|\bar p_0,\sigma) &=& \int dp_0 p_0G(p_0|\bar p_0,\sigma)
    \mathrm{e}^{-ip_0 t} F_i(z|p_0)\nonumber\\
    &=& \bar p_0\Phi_i(t,z|\bar p_0,\sigma) + \int dp_0 (p_0-\bar p_0)G(p_0|\bar p_0,\sigma)\mathrm{e}^{-ip_0t}F_i(z|p_0).
\end{eqnarray}

Considering
\begin{equation}
    (p_0-\bar p_0)G(p_0|\bar p_0,\sigma) = \sigma^2\frac{\partial}{\partial\bar p_0}G(p_0|\bar p_0,\sigma),
\label{eq:Gaussian_center_derivative}
\end{equation}
we have
\begin{eqnarray}
 i\frac{\partial}{\partial t}\Phi_i(t,z|\bar p_0,\sigma) &=& \left[e E zI_2+M_n\sigma_x-i\sigma_z\partial_z\right]\Phi_i(t,z|\bar p_0,\sigma)\nonumber\\
 &=& \left(\bar p_0 + \sigma^2\frac{\partial}{\partial\bar p_0}\right)\Phi_i(t,z|\bar p_0,\sigma).
\label{eq:finite_width_packet_equation}
\end{eqnarray}

We now introduce the Fourier transform of the packet $\Phi_i(t,z|\bar p_0,\sigma)$,
\begin{equation}
    \widetilde\Phi_i(t,p_z|\bar p_0,\sigma) = \frac{1}{\sqrt{2\pi}}\int_{-\infty}^{+\infty}dz \mathrm{e}^{-ip_z z}\Phi_i(t,z|\bar p_0,\sigma),
\label{eq:packet_Fourier_transform}
\end{equation}
Eq.~\eqref{eq:finite_width_packet_equation} means
\begin{equation}
    \left[ie E\frac{\partial}{\partial p_z} + p_z\sigma_z + M_n\sigma_x\right]\widetilde\Phi_i(t,p_z|\bar p_0,\sigma) =
    \left(\bar p_0 + \sigma^2\frac{\partial}{\partial\bar p_0}\right)\widetilde\Phi_i(t,p_z|\bar p_0,\sigma),
\label{eq:momentum_packet_equation}
\end{equation}
and Eq.~\eqref{eq:translation_property_packet2} implies
\begin{equation}
    \widetilde\Phi_i(t,p_z|\bar p_0,\sigma) = \mathrm{e}^{-i\bar p_0 Q_t}\widetilde\Phi_i(t,p_z|0,\sigma)
\end{equation}
with $Q_t(p_z)=t+p_z/(eE)$.

The Gaussian packet obeys the convolution semigroup relation
\begin{equation}
    G(p_0|\bar p_0,\sigma_2) = \int_{-\infty}^{+\infty}dq G(q|\bar p_0,\tau)G(p_0|q,\sigma_1), \qquad \sigma_2^2=\sigma_1^2+\tau^2
\label{eq:Gaussian_semigroup}
\end{equation}
which results in 
\begin{eqnarray}
    \widetilde\Phi_i(t,p_z|\bar p_0,\sigma_2) &=& \int_{-\infty}^{+\infty}dq G(q|\bar p_0,\tau)\widetilde\Phi_i(t,p_z|q,\sigma_1)\nonumber\\
    &=& \widetilde\Phi_i(t,p_z|0,\sigma_1)\int_{-\infty}^{+\infty}dq G(q|\bar p_0,\tau)\mathrm{e}^{-iqQ_t}\nonumber\\
    &=& \mathrm{e}^{-\tau^2Q_t^2/2}\widetilde\Phi_i(t,p_z|\bar p_0,\sigma_1).
\end{eqnarray}
Therefore the quantity
\begin{equation}
    f_i(t,p_z) = \mathrm{e}^{\sigma^2Q_t^2/2}\widetilde\Phi_i(t,p_z|0,\sigma)
\label{eq:reduced_f_packet_definition_time}
\end{equation}
is independent of the packet width $\sigma$, and we have
\begin{equation}
    \widetilde\Phi_i(t,p_z|\bar p_0,\sigma) = \mathrm{e}^{-i\bar p_0 Q_t} \mathrm{e}^{-\sigma^2Q_t^2/2}f_i(t,p_z).
\label{eq:packet_factorization_before_time}
\end{equation}
Substituting Eq.~\eqref{eq:packet_factorization_before_time} into Eq.~\eqref{eq:momentum_packet_equation} gives
\begin{equation}
    ie E \frac{\partial f_i(t,p_z)}{\partial p_z} + \left(p_z\sigma_z + M_n\sigma_x\right)f_i(t,p_z) = 0.
\label{f_equation}
\end{equation}
On the other hand, the Fourier transformation of Eq.~\eqref{eq:finite_width_packet_equation} leads to 
\begin{equation}
    i\frac{\partial}{\partial t}\widetilde\Phi_i(t,p_z|\bar p_0,\sigma) = \left[ie E\frac{\partial}{\partial p_z} + p_z\sigma_z
        + M_n\sigma_x\right]\widetilde\Phi_i(t,p_z|\bar p_0,\sigma)
\label{eq:packet_time_evolution_pz}
\end{equation}
which shows 
\begin{equation}
    \frac{\partial f_i(t,p_z)}{\partial t}=0.
\label{eq:f_time_independence}
\end{equation}
Therefore, we can neglect the time dependence and write
\begin{equation}
    f_i(p_z) = f_i(t,p_z).
\end{equation}

Since $p_z\sigma_z+M_n\sigma_x$ is Hermitian, we have
\begin{eqnarray}
    \frac{d}{dp_z}\left(f_i^\dagger f_j \right) &=& -\frac{i}{e E}f_i^\dagger \left(p_z\sigma_z+M_n\sigma_x\right)f_j +
    \frac{i}{e E}f_i^\dagger\left(p_z\sigma_z+M_n\sigma_x\right)f_j\nonumber\\
    &=& 0. 
\end{eqnarray}
This means that, for any two solutions $f_i(p_z)$ and $f_j(p_z)$, their product $f_i^\dagger(p_z)f_j(p_z)$ is independent of $p_z$ and can be evaluated at any convenient value of $p_z$. Evaluating it at $p_z=0$, we obtain
\begin{eqnarray}
    f_1^\dagger(p_z)f_1(p_z) &=& 2\mathrm{e}^{3\pi\mu_n/4},\nonumber\\
    f_4^\dagger(p_z)f_4(p_z) &=& 2\mathrm{e}^{3\pi\mu_n/4}\left(\mathrm{e}^{\pi\mu_n}-1\right),\nonumber\\
    f_1^\dagger(p_z)f_4(p_z) &=& 0.
\end{eqnarray}

We next consider the normalized time-dependent longitudinal packets
\begin{equation}
    \mathcal U_i(t,z|\bar p_0,\sigma) = C_i \Phi_i(t,z|\bar p_0,\sigma),
\label{eq:normalized_longitudinal_packet}
\end{equation}
with the normalization constants
\begin{equation}
    C_1= \frac{\mathrm{e}^{-3\pi\mu_n/8}}{\sqrt{4\pi eE}}, \qquad C_4 = \frac{\mathrm{e}^{-3\pi\mu_n/8}}{\sqrt{4\pi e E}\sqrt{\mathrm{e}^{\pi\mu_n}-1}},
\end{equation}
their momentum-space representation is
\begin{equation}
    \widetilde {\mathcal{U}}_i(t,p_z|\bar p_0,\sigma) = C_i \mathrm{e}^{-i\bar p_0 \left(t+p_z/(eE)\right)}\mathrm{e}^{-\sigma^2\left(t+p_z/(eE)\right)^2/2}
    f_i(p_z).
\label{eq:normalized_longitudinal_packet_pz}
\end{equation}
For arbitrary finite widths $\sigma,\tau>0$, Plancherel's theorem yields, 
\begin{eqnarray}
    \int_{-\infty}^{+\infty} dz \mathcal U_i^\dagger(t,z|\bar p_0,\sigma)\mathcal U_j(t,z|\bar p_0',\tau)  
    &=& C_iC_j\int_{-\infty}^{+\infty}dp_z \mathrm{e}^{i(\bar p_0-\bar p_0')(t+p_z/(eE))}\mathrm{e}^{-(\sigma^2+\tau^2)(t+p_z/(eE))^2/2}
    f_i^\dagger(p_z)f_j(p_z)\nonumber\\
    &=& \frac{\delta_{ij}}{2\pi e E}\int_{-\infty}^{+\infty}dp_z \mathrm{e}^{i(\bar p_0-\bar p_0')(t+p_z/(eE))}\mathrm{e}^{-(\sigma^2+\tau^2)
    (t+p_z/(eE))^2/2}\nonumber\\
    &=& {\delta_{ij}\over \sqrt{2\pi(\sigma^2+\tau^2)}}\mathrm{e}^{-{(\bar p_0-\bar p_0')^2\over 2(\sigma^2+\tau^2)}}.
\label{eq:time_packet_overlap_before_shift}
\end{eqnarray}
Therefore, in the zero-width limit, we have
\begin{equation}
    \lim_{\sigma,\tau\to 0^+}\int_{-\infty}^{+\infty}dz \mathcal U_i^\dagger(t,z|\bar p_0,\sigma)\mathcal U_j(t,z|\bar p_0',\tau)
    = \delta_{ij} \delta(\bar p_0-\bar p_0').
\label{eq:generalized-longitudinal-orthogonality}
\end{equation}
Using Eqs.~\eqref{eq:UV_orth_packet} and \eqref{eq:generalized-longitudinal-orthogonality}, we obtain finally the orthonormality relation (\ref{ortho}) for the wave functions in the in-mode basis $\{\psi^{(1)},\psi^{(4)}\}$. For the out-mode basis $\{\psi^{(3)},\psi^{(2)}\}$, the similar relations can be obtained from the unitarity of the Bogoliubov transformation.

\subsection{Completeness}
We now proceed to prove the completeness relation. Using Eq.~\eqref{eq:normalized_longitudinal_packet_pz}, we have 
\begin{eqnarray}
    \sum_{i\in\{1,4\}}\int_{-\infty}^{+\infty}d\bar p_0 \widetilde{\mathcal{U}}_i(t,p_z|\bar p_0,\sigma)\widetilde{\mathcal{U}}_i^\dagger(t,p_z'|\bar p_0,\sigma) &=& \mathrm{e}^{-{\sigma^2}
    \left[Q_t(p_z)^2+Q_t(p_z')^2\right]/2}\nonumber\\
    && \times \int_{-\infty}^{+\infty}d\bar p_0\, \mathrm{e}^{-i\bar p_0 \left[Q_t(p_z)-Q_t(p_z')\right]}
    \sum_{i\in\{1,4\}}C_{i}^2 f_i(p_z)f^{\dagger}_i(p_z')\nonumber\\
    &=& \mathrm{e}^{-\sigma^2(t+p_z/(eE))^2}\delta(p_z-p_z')I_2
\label{com1}
\end{eqnarray}
for the longitudinal sector, its Fourier transformation gives
\begin{equation}
    \sum_{i\in\{1,4\}} \int_{-\infty}^{+\infty}d\bar p_0 \mathcal U_i(t,z|\bar p_0,\sigma)\mathcal U_i^\dagger(t,z'|\bar p_0,\sigma)
    = \mathrm{e}^{-ie E t(z-z')}\frac{e E}{2\sqrt{\pi}\sigma}\mathrm{e}^{-{(e E)^2(z-z')^2\over 4\sigma^2}}I_2.
\label{eq:K_z_resolution_time}
\end{equation}
In the zero-width limit, it becomes
\begin{equation}
    \lim_{\sigma\to 0^+}\sum_{i\in\{1,4\}} \int_{-\infty}^{+\infty}d\bar p_0 \mathcal U_i(t,z|\bar p_0,\sigma)\mathcal U_i^\dagger(t,z'|\bar p_0,\sigma) = \delta(z-z')I_2.
\label{eq:longitudinal_completeness_final}
\end{equation}

For the transverse sector, using the harmonic-oscillator completeness relation~\eqref{eq:harmonic-oscillator-completeness}, we obtain
\begin{eqnarray}
&&    \sum_{sn}U_{sn}(x)U_{sn}^\dagger(x') = \delta(\xi_1-\xi_1')(\chi_{++}\chi_{++}^\dagger + \chi_{+-}\chi_{+-}^\dagger),\nonumber\\
&&    \sum_{sn}V_{sn}(x)V_{sn}^\dagger(x')/M_n^2 = \delta(\xi_1-\xi_1')(\chi_{-+}\chi_{-+}^\dagger + \chi_{--}\chi_{--}^\dagger).
\label{eq:V_transverse_completeness}
\end{eqnarray}
Combining Eqs.~\eqref{eq:longitudinal_completeness_final} and \eqref{eq:V_transverse_completeness} gives the completeness relation (\ref{complete}) in the in-mode basis. The unitarity of the Bogoliubov transformation leads to the similar relation for the out-mode basis.

\bibliographystyle{apsrev4-2}
\bibliography{ref}

\end{document}